\documentclass[11pt,letterpaper]{article}

\usepackage[margin=1in]{geometry}
\usepackage{latexsym,graphicx,amssymb}
\usepackage{amsthm}
\usepackage{amsmath,enumerate}
\usepackage{wrapfig}
\usepackage{subfigure}
\usepackage{float}
\usepackage{xspace}
\usepackage{paralist}
\usepackage{enumerate} 
\usepackage{cases}
\usepackage{caption}
\usepackage{tabularx}
\usepackage{algpseudocode}
\usepackage[linesnumbered,lined,boxed,commentsnumbered]{algorithm2e}
\usepackage{color}
\usepackage{complexity}

\usepackage{times}

\usepackage{multicol}

\newcounter{note}[section]

\newcommand{\yuetodo}[1]{{\large\color{green}[Yue todo: #1]}}

\definecolor{yue}{rgb}{0.7, 0, 0}

\renewcommand{\yuetodo}[1]{}

\newcommand{\mcal}[1]{\ensuremath{\mathcal {#1}}}

\newcommand{\ceil}[1]{\ensuremath{\left \lceil #1 \right \rceil}}

\definecolor{darkgreen}{rgb}{0,0.5,0}
\definecolor{lightblue}{RGB}{0,176,240}
\definecolor{darkblue}{RGB}{0,112,192}
\definecolor{lightpurple}{RGB}{124, 66, 168}
\definecolor{grey}{RGB}{139, 137, 137}
\definecolor{maroon}{RGB}{178, 34, 34}
\definecolor{green}{RGB}{34, 139, 34}
\definecolor{types}{RGB}{72, 61, 139}
\definecolor{gold}{rgb}{0.8, 0.33, 0.0}

\definecolor{darkgray}{gray}{0.3}

\definecolor{darkred}{rgb}{0.5, 0, 0}
\definecolor{darkgreen}{rgb}{0, 0.5, 0}
\definecolor{darkblue}{rgb}{0,0,0.5}

\newcommand\markx[2]{}

\newcommand{\Z}{\mathbb{Z}}

\newcommand{\Adv}{\ensuremath{\mathsf{Adv}}\xspace}

\newcommand{\ignore}[1]{}

\newcommand{\conv}{\mathsf{conv}}
\newcommand{\coA}{\conv(\mcal{A})}

\renewcommand{\RP}{\ensuremath{\mathsf{RP}}\xspace}
\newcommand{\PS}{\ensuremath{\mathsf{PS}}\xspace}
\renewcommand{\PP}{\ensuremath{\mathsf{PP}}\xspace}

\newcounter{task}

\newtheorem{theorem}{Theorem}[section]

{
\theoremstyle{definition}
\newtheorem{remark}[theorem]{Remark}
}

\newtheorem{corollary}[theorem]{Corollary}
\newtheorem{fact}[theorem]{Fact}
\newtheorem{lemma}[theorem]{Lemma}

{
\theoremstyle{definition}
\newtheorem{definition}[theorem]{Definition}
}

{
\theoremstyle{definition}
\newtheorem{example}[theorem]{Example}
}

\newcounter{cnt:challenge}

\newcommand*\samethanks[1][\value{footnote}]{\footnotemark[#1]}

\SetKwComment{Comment}{//}{}

\numberwithin{figure}{section}
\numberwithin{equation}{section}

\title{Game-Theoretically Secure Protocols for the Ordinal Random Assignment Problem}

\author{T-H. Hubert Chan\thanks{Department of Computer Science, The University of Hong Kong. {\texttt{hubert@cs.hku.hk, twen.hku@gmail.com, hxie@connect.hku.hk, 
csxuequan@connect.hku.hk}}} \and 
Ting Wen\samethanks \and
Hao Xie\samethanks \and
Quan Xue\samethanks}

\date{}

\begin{document}

\begin{titlepage}

\maketitle

\begin{abstract}
We study game-theoretically secure protocols for the classical ordinal assignment problem (aka matching with one-sided preference), in which each player has a total preference order on items.  To achieve the fairness notion of equal treatment of equals, conventionally the randomness necessary to resolve conflicts between players is assumed to be generated by some trusted authority.  However, in a distributed setting, the mutually untrusted players are responsible for generating the randomness themselves.

In addition to standard desirable properties such as fairness and Pareto-efficiency, we investigate the game-theoretic notion of maximin security, which guarantees that an honest player following a protocol will not be harmed even if corrupted players deviate from the protocol. Our main contribution is an impossibility result that shows no maximin secure protocol can achieve both fairness and ordinal efficiency.  Specifically, this implies that the well-known probabilistic serial (PS) mechanism by Bogomolnaia and Moulin cannot be realized by any maximin secure protocol.

On the other hand, we give a maximin secure protocol that achieves fairness and stability (aka ex-post Pareto-efficiency).  Moreover, inspired by the PS mechanism, we show that a variant known as the OnlinePSVar (varying rates) protocol can achieve fairness, stability and uniform dominance, which means that an honest player is guaranteed to receive an item distribution that is at least as good as a uniformly random item.  In some sense, this is the best one can hope for in the case when all players have the same preference order.

\noindent\textbf{keywords: }{Ordinal assignment problem \and Distributed protocols  \and Game-theoretic security}
\end{abstract}

\thispagestyle{empty}
\end{titlepage}

\section{Introduction}
\label{sec:intro}

In this paper, we study secure distributed protocols for
the classical \emph{ordinal assignment problem}~\cite{Gardenfors1973,HyllandZeckhauser79,BogoMoulin2001} (aka \emph{matching with one-sided preference}), in which there
are players~$\mcal{N}$ and items~$\mcal{M}$,
where each player has some total preference order on the items.
For ease of illustration, we focus on the case $n=|\mcal{N}| = |\mcal{M}|$, 
even though our results can be readily generalized to the case $|\mcal{N}| \neq |\mcal{M}|$.

A \emph{mechanism} takes a preference profile of all players' preference orders
and returns a (possibly random) assignment of items
to players, where an assignment is a matching between 
$\mcal{M}$ and $\mcal{N}$, i.e., a bipartite graph in
$\mcal{M} \times \mcal{N}$, where the degree of each node is at most 1.

From a player's perspective, the result of the mechanism
is the probability vector representing the distribution 
of the item that it receives. A player's preference
naturally induces a partial preference order on the probability vectors.  We assume that a player
prefers to receive some item over having no item.

In the economics literature, several mechanism properties have been investigated.
\begin{compactitem}
\item \emph{Pareto Efficiency.}  Intuitively, this means a mechanism attempts to
cater to the preference orders of the players.  

An assignment
is \emph{stable} (aka \emph{ex-post} Pareto-efficient) if
there is no subset $S$ of players who would like to exchange 
items such that everyone in $S$ gets a more preferred item afterwards;
a mechanism is stable if it always returns a stable assignment.

A (random) assignment is \emph{ordinally efficient} (aka \emph{ex-ante} Pareto-efficient)
if there does not exist another random assignment such that a non-empty subset $S$ of players 
receive item probability vectors they strictly prefer than before (while those for players not in $S$ do not change).
A mechanism is ordinally efficient if for any player preference profile,
it returns an ordinal efficient assignment.
One could see that ordinal efficiency is a stronger property than stability.

\item \emph{Fairness.}  This is also known as
equal treatment (of equals), meaning that if two players
have an identical preference order, then under the mechanism,
the two players should receive identical item distributions.

In this paper, we will consider a stronger notion of fairness
that places conditions when two players have an identical preference
among a subset of their most preferred items.

\item \emph{Truthfulness.} This is also known as \emph{strategyproof},
which means a player does not have the incentive to misreport
their preference order to a mechanism.
\end{compactitem}

\noindent \textbf{Distributed Randomness to Achieve Equal Treatment.}  Observe that
randomness is necessary for a mechanism to achieve equal treatment.
Typically, in the economics literature~\cite{BogoMoulin2001}, one assumes that
some trusted central authority will be responsible for generating the randomness in a mechanism,
and it suffices to analyze a mechanism as a function
that takes a preference profile and returns a distribution of assignments.
In contrast, in a distributed setting
such as blockchain applications~\cite{nakamoto2009bitcoin},
there is no trusted authority and any randomness 
is generated in a distributed fashion among the players,
each of whom may want to receive a more preferable item distribution
or behave maliciously to harm other players.
Hence, we will explore various security notions for distributed protocols.

\noindent \textbf{Model of Distributed Protocols.}
We consider the following assumptions that are commonly adopted
in blockchain applications: 
(i) the protocol is distributed and involves only the players (with no trusted authority), (ii) each message can be seen by everyone.
Specifically,
we consider a synchronized communication model,
in which each player can post messages to some broadcast channel (such as
a \emph{ledger}~\cite{DBLP:conf/crypto/BadertscherMTZ17}). 
In each round,
each agent reads posted messages on the channel from previous rounds,
performs some local computation (possibly based on locally generated randomness) and posts new messages to the channel.
At the end of the protocol,
some publicly agreed deterministic function is applied to the whole transcript of messages
to identify the output.

An \emph{honest} agent follows the procedure as specified by the protocol.
To distinguish between truthfulness and honesty, we assume
that the preference profile is either publicly known
or each player has already declared some preference order before the protocol begins.

In this paper, we also distinguish between mechanism and protocol in the following sense.
We say that a protocol \emph{realizes} a mechanism, if under honest execution by all players,
the protocol produces a random assignment that has the same
distribution as specified by the mechanism.  When we say that
a protocol has a certain mechanism property (such as equal treatment or ordinal efficiency),
we mean that the property is satisfied when all players behave honestly in the protocol.

On the contrary, an \emph{adversary} controls some \emph{corrupted} players
that may deviate from the protocol.  A \emph{Byzantine} adversary can cause a corrupted
player to behave arbitrarily, while a \emph{fail-stop} adversary can only cause a
corrupted player to abort (i.e., stop sending messages) in a protocol.

\noindent \textbf{Security Notions of Distributed Protocols.} 
The strictest notion of security for a protocol solving
the problem is that under any strategy of the adversary,
the output of the protocol still has the same distribution
as one under honest execution.
However, this is impossible even for the simple
case of the 
\emph{fair coin toss} problem~\cite{DBLP:conf/stoc/Cleve86,DBLP:conf/tcc/ChungGLPS18},
in which $2$ players wish to agree on a uniformly random bit in $\{0,1\}$ (with
zero bias).  To see that this is a special case
of the assignment problem, consider 2 players that have the same
preference order on 2 items.  Then, any stable mechanism that achieves
equal treatment is equivalent to returning one of the 2 possible
assignments with equal probability.
Specifically, Cleve's impossibility result~\cite{DBLP:conf/stoc/Cleve86} states that
given any protocol for the fair coin toss problem
among two players that terminates within a bounded number of rounds, 
at least one of the players can cause the output to have a non-zero
probability bias towards either 0 or 1 by aborting at some point during the protocol.
This impossibility result holds even if one assumes ideal cryptographic primitives
such as one-way functions.

\noindent \textbf{Game Theoretic Notions of Security.}
Observe that Cleve's aforementioned impossibility
result states that a player can bias the outcome of the protocol,
but not necessarily towards a more favorable one to itself.
As opposed to the fair coin problem (in which the goal
of the adversary is to introduce bias),
in the \emph{lottery} problem~\cite{DBLP:conf/eurosp/MillerB17} (aka \emph{leader election problem}),
exactly one of the $n$ players is chosen as the \emph{winner}.
For the lottery problem, a protocol is \emph{maximin} secure~\cite{DBLP:conf/crypto/ChungCWS21} if an honest player's 
winning probability does not decrease under the strategy
of an adversary.

The lottery problem can be solved by an elegant distributed protocol
with the help of a non-malleable commitment scheme (e.g., one
based on one-way functions~\cite{DBLP:journals/jacm/LinP15}).
Intuitively, such a scheme allows an agent~$i$ to hide some input $x_i$ in a commitment~$C_i$,
which behaves like a blackbox to others;
later, the agent can decide to open $C_i$ to reveal $x_i$,
but the scheme prevents $C_i$ from opening to any other different value.
To simplify our description, we assume the existence of an ideal commitment scheme;
this has the advantage of separating the computational issue regarding
cryptography from the game theoretic aspects of the problem.

The special case of $n=2$ agents can be solved by
a simple Blum duel protocol~\cite{DBLP:journals/tocs/Blum83},
in which each of two players (labeled 0 and 1) randomly picks an input bit in $\{0,1\}$ 
and broadcasts its commitment. After receiving another player's commitment,
each player opens its own commitment to reveal its input bit,
and the winner is the XOR of the two revealed input bits.  However, if one player does
not open its input bit, then the other player will automatically be the winner of the duel.
Observe that an honest player wins with a probability of at least $\frac{1}{2}$.
(Since this is a zero sum game, no dishonest player can win with a probability of larger than $\frac{1}{2}$.)
Using a binary tournament tree of depth $O(\log n)$ in which every internal node corresponds to an instance of the duel subroutine, one can see that this readily corresponds to a distributed protocol with $O(\log n)$ rounds for the lottery problem in which an honest player wins with a probability of at least $\frac{1}{n}$.

Motivated by the lottery problem, the notion of maximin security 
can also be applied to a protocol for the assignment problem,
in which an honest player would receive  the same or a more preferable
outcome distribution, should corrupted players deviate
from the protocol.

\subsection{Technical Challenges}
\label{sec:tech}

To understand this game-theoretic notion of security, we first investigate whether
well-known mechanisms in the literature can
be realized by maximin secure protocols.

\noindent \textbf{Random Priority} ($\mathsf{RP}$ aka \emph{random serial dictatorship} ) mechanism~\cite{Zhou90,RandomPriority}. The mechanism
first samples a uniformly random permutation on the players,
who are assigned items sequentially, one player at a time accordingly.
When it is a player's turn,
 it will receive its most preferred item among the
still available items.  It can be easily checked that
the mechanism is truthful and achieves equal treatment,
but it is known to be not ordinally efficient.

Observe that to realize $\mathsf{RP}$,
one possible approach is to generate a permutation
uniformly at random in a ``maximin secure'' fashion.
Since we already have a maximin secure protocol
for the lottery problem, it is tempting to use it to
generate a random permutation of players.  For instance,
an instance of the lottery problem can determine which player ranks first,
and so on for the rest of the permutation.
Indeed, one can show that this protocol is maximin secure
with respect to the rank received by a player in the permutation.

Unfortunately, this does not translate to the maximin security
with respect to a player's preference for items.  Consider
the following example with 3 players such that players~1 and~2
both have item~$A$ as their favorite, while player~3 has a different
favorite item~$B$.  Observe that under honest execution,
player~1 receives its favorite item~$A$ with probability $\frac{1}{2}$.

However, player~1 can be hurt in the following way.
When rank 1 is determined in the first lottery problem instance,
player~3 has an abort strategy that it aborts whenever in the round against player~2,
which results in player~2 winning automatically in this round.
Under this strategy of player~3, player~1 still wins with 
a probability~$\frac{1}{3}$, but the winning probability of player~3
can be transferred to player~2 who now wins with probability~$\frac{2}{3}$.
As a result, under this attack, player~1 receives
its favorite item with probability~$\frac{1}{3}$, which is smaller than before.

The above example shows that generating a random
permutation via the lottery problem protocol cannot
achieve maximin security for the assignment problem,
but does not rule out the possibility that there
may be a maximin secure protocol that can realize \RP.

\noindent \textbf{Probabilistic serial} (\PS) mechanism.
This was proposed
by Bogomolnaia and Moulin~\cite{DBLP:journals/jet/BogomolnaiaM01}, for which
we imagine that each object is one unit of different juice,
and each player consumes its most preferred available juice 
in the order of its preference list
at the same rate; the resulting consumption corresponds to a (deterministic) fractional assignment of the object
which is a bistochastic matrix 
\footnote{A bistochastic matrix is one with non-negative real elements such that the sum of every row and the sum of every column is equal to $1$.}
 that can be random rounded to give an (integral) assignment.
While it is known that \PS is ordinally efficient and clearly achieves equal treatment,
there are known examples in which \PS is not truthful.
Indeed, it has been proved~\cite{DBLP:journals/jet/BogomolnaiaM01}
that in general, no mechanism can simultaneously achieve equal treatment,
truthfulness and ordinal efficiency.
Realizing \PS by a maximin secure protocol seems tricky, because the rounding
of the aforementioned bistochastic matrix involves intricate dependencies
of item preferences among the players. Indeed,
our main result shows that this is actually impossible.

\subsection{Our Contributions}

Analogous to the aforementioned impossibility result~\cite{DBLP:journals/jet/BogomolnaiaM01}
that no mechanism can satisfy equal treatment, truthfulness and ordinal efficiency simultaneously,
we have the following impossibility result for maximin secure protocols.

\begin{theorem}[Impossibility Result to Achieve Maximin Security]
\label{th:main_impossible}
For $n \geq 4$ players, any mechanism that achieves both
strong equal treatment and ordinal efficiency cannot be realized
by a maximin secure protocol (against a fail-stop adversary) 
that terminates with a bounded number of rounds.
\end{theorem}

Here are the implications of this impossibility result.

\begin{compactitem}
\item Strong equal treatment means that 
for any $k \leq n$,
if two players have
exactly the same preference order among their
$k$ most favorite items, 
then the two players have exactly the same probabilities
for receiving each of those $k$ items.
Since \PS also achieves strong equal treatment,
it follows that no maximin secure protocol can realize \PS.

\item  The impossibility
result also means that if a protocol ensures
that an honest player will not be hurt by corrupted players 
(i.e., maximin security is achieved),
then the mechanism is not ordinally efficient,
which implies that it is possible that all the players 
might collude and deviate from the protocol such
that no player will get hurt and some player will be strictly better off.
\end{compactitem}

Even though we do not know how to realize \RP
with a maximin secure protocol and
have shown that \PS cannot be realized
by a maximin secure protocol (against even a fail-stop adversary),
we have the following positive result on maximin secure protocols.

\begin{theorem}[Maximin Secure Protocol]
\label{th:main_maximin}
There exists a mechanism that achieves both
stability and strong equal treatment (when
all players are honest) and can
be realized by a maximin secure protocol against a fail-stop adversary
controlling up to $n-1$ corrupted players.
\end{theorem}

Our protocol known as \emph{preference priority} (\PP, Algorithm~\ref{alg:PP})
runs a sequence of lottery problem instances,
where each lottery decides the fate of a specific item.
Loosely speaking, the protocol can achieve maximin security
because it ensures that a fail-stop adversary cannot affect the order of
the lottery problem instances in the sequence.
In Section~\ref{sec:prelim}, we explain some scenarios
in which it is justifiable to consider only fail-stop adversaries.

\noindent \textbf{Uniform Dominance.}  The notion of maximin security
guarantees that an honest player cannot be hurt by
corrupted players that deviate from the protocol,
but an honest player can still be attacked if
other players lie about their preference orders.
If a mechanism satisfies equal treatment,
then an honest player can be attacked by a malicious adversary that controls every other player, who claims to have exactly the same preference order as the honest player, thereby forcing everyone to receive every item with the same probability.
Therefore,
the adversary can make sure that an honest player
cannot get something better than a uniformly random item in $\mcal{M}$.
We say that a protocol achieves \emph{uniform dominance}
if an honest player receives an item distribution that is at least
as good as a uniformly random item, no matter what the other players say (about
their preference orders) or do (in the protocol).

It is not too difficult to check that the above idea of
realizing \RP by generating a uniformly random permutation
via instances of the lottery problem can achieve uniform dominance.
On the other hand, the \PS mechanism ensures that for any preference profile,
each player receives an item distribution that 
is at least as good as a uniformly random item.
Even though we have shown that \PS cannot be realized by a maximin secure
protocol, we have designed a variant known as \emph{online \PS with varying rates}
($\mathsf{OnlinePSVar}$, Algorithm~\ref{alg:onlineps}) that achieves uniform dominance.

\begin{theorem}[Uniform Dominance]
\label{th:main_uniform}
The $\mathsf{OnlinePSVar}$ protocol achieves
stability, strong equal treatment and
uniform dominance against a Byzantine adversary (controlling
up to $n-1$ players).
\end{theorem}

Even though $\mathsf{OnlinePSVar}$ also
uses the lottery problem subroutine,
it might have a potential advantage over \RP
when players have vastly different preference orders.
For \RP, observe that all players need to participate
in the lottery problem to determine which player has ranked 1 in the permutation.
On the other hand, one can check that for $\mathsf{OnlinePSVar}$,
when players have very different favorite items,
each instance of the lottery problem can potentially involve
fewer players (because each item might be fractionally consumed by fewer players), thereby improving the round complexity
of the protocol, as the lottery problem on $n$ players
takes $O(\log n)$ rounds.

\noindent \textbf{Paper Organization.}  We give the formal notation in Section~\ref{sec:prelim}
and introduce standard building blocks in Section~\ref{section:blocks}.
Our impossibility result in Theorem~\ref{th:main_impossible}
is proved in Section~\ref{section:impossible}.
Our maximin secure protocol is given in Section~\ref{sec:algo},
and we show how uniform dominance is achieved in Section~\ref{sec:uniform_dom}.
Finally, we outline some future directions in Section~\ref{sec:conclusion}.
All omitted proofs are given in the Supplementary Materials.

\subsection{Other Related Work}
\label{sec:related}

Since the ordinal assignment problem
was introduced by Gardenfors~\cite{Gardenfors1973},
there have been numerous works on the subject;
for details, refer to Chapter~2 of the book~\cite{onlinematchingbook}.

To circumvent Cleve's aforementioned impossibility result~\cite{DBLP:conf/stoc/Cleve86}
for the fair coin toss problem (with multi-players), Chung et al.~\cite{DBLP:conf/tcc/ChungGLPS18}
have proposed game-theoretic notions of security when
players have a preference for the coin outcome.

The folklore tournament tree protocol for the lottery problem
has gained renewed interest in the context of blockchain applications~\cite{DBLP:conf/eurosp/MillerB17}.  To improve the round complexity of lottery protocols,
Chung et al.~\cite{DBLP:conf/crypto/ChungCWS21}
have considered approximate game-theoretic notions of security.
Since the lottery problem has a clear zero-sum game structure,
an honest player cannot be hurt \emph{iff} corrupted players
cannot gain any unfair advantage.  In contrast,
our impossibility result in Theorem~\ref{th:main_impossible} for the ordinal assignment problem
implies that if an honest player cannot be hurt in a protocol,
then it might still be possible for some players to collude 
and be strictly better off.

\section{Preliminaries}
\label{sec:prelim}

Let $\mcal{N}$ denote the set of players
and $\mcal{M}$ denote the set of items,
where $n = |\mcal{N}| = |\mcal{M}|$. 
For a positive integer~$\ell$,
we write $[\ell] := \{1, 2, \ldots, \ell\}$.
We use $\mcal{S}_{\mcal{M}}$ (or $\mcal{S}$ when $\mcal{M}$ is clear from context) to denote the collection of
total orders over $\mcal{M}$.
Each player $i \in \mcal{N}$ has some preference order $\succ_i$ in $\mcal{S}$, 
which is also represented
by a \emph{favorite} function $O_i: [n] \rightarrow \mcal{M}$,
where $O_i(k)$ is the $k$-th favorite item of player~$i$.
Given a preference profile $\sigma \in \mcal{S}^{\mcal{N}}$,
we implicitly assume that the associated
$\succ_i$ (also denoted as $\sigma_i$) and $O_i$ are defined for each $i \in \mcal{N}$.

We use $\mcal{A}$ to denote
the collection of \emph{assignment} matrices in $\{0,1\}^{\mcal{N} \times \mcal{M}}$
such that every row and column has exactly one non-zero entry.
For instance, given some $P \in \mcal{A}$, $P(i,j) = 1$ \emph{iff} player~$i$
receives item~$j$; we also use $P_i$ to denote the $i$-th row of $P$.
The convex hull\footnote{A convex hull of $S\subset \R^n$ refers to the minimum convex set that contains~$S$.}~$\coA=\{P \in [0,1]^{\mcal{N} \times \mcal{M}}: \sum_{j\in \mcal{M}} P(i',j)=1, \sum_{i\in \mcal{N}} P(i,j')=1, \forall i' \in \mcal{N}, j'\in\mcal{M}\}$
is exactly the collection of bistochastic matrices.

\noindent \textbf{Distribution.}
Given some set $\mcal{U}$,
we use $\Delta(\mcal{U}) := \{x \in [0,1]^{\mcal{U}}:
\sum_{u \in \mcal{U}} x_u = 1 \}$
to denote the collection of distributions on $\mcal{U}$.

\noindent \textbf{Mechanism.}  In this paper,
a \emph{mechanism} is a mapping that takes a
preference profile in  $\mcal{S}^{\mcal{N}}$ 
and returns a distribution in
$\Delta(\mcal{A})$.  Typically, the description
of a mechanism gives a method to randomly sample
an assignment in $\mcal{A}$.
Observe that a distribution $\rho \in \Delta(\mcal{A})$ induces
a bistochastic matrix $\sum_{A \in \mcal{A}} \rho_A \cdot A \in \coA$.  Alternatively, in the literature,
a mechanism is sometimes described by giving the resulting
bistochastic matrix, from which a (possibly non-unique) distribution
of assignments can be computed efficiently.
However, note that it can be NP-hard to compute
the bistochastic matrix from 
a mechanism description (such as \RP~\cite{SabanS2015}).

Conventionally, the randomness used for sampling an assignment
in a mechanism is assumed to be generated by some trusted authority.
\emph{Truthfulness} refers to whether a player reveals its true preference
order to the mechanism.  
The main focus of this work is the
scenario when this randomness
is jointly generated by the players according to some procedure
known as a (distributed) \emph{protocol}.

\noindent \textbf{Communication Model of Protocols.}
Players participate
in a (possibly randomized) protocol, at the end of which the whole transcript
of all sent messages determines an assignment in $\mcal{A}$.  
We assume that either the preference profile is public information, or 
before the protocol begins, each player declares its preference order.
We emphasize the distinction that \emph{honesty} refers to whether a player follows the procedure as specified by the protocol,
as opposed to whether a player is truthful about its preference.

A protocol proceeds in synchronous rounds
over a broadcast channel, i.e., a message sent 
by a player in one round will reach all players at the beginning of the
next round.
In every round,
based on messages received in previous rounds,
a player generates randomness and performs local computation
as specified by the protocol to generate a message to be sent
in this round.

\noindent \textbf{Adversarial Model.} An adversary \Adv controls some \emph{corrupted} players.
The adversary can observe the internal states of the corrupted players
and control their actions.
We assume that the adversary is \emph{rushing}, i.e.,
it can
wait for the messages from all honest players in a round before
it decides the actions of the corrupted players in that round.
A \emph{fail-stop} adversary can instruct a corrupted player to deviate
from the protocol only by stopping to broadcast a message in some round
(after which the player will not broadcast any message in subsequent rounds).
A Byzantine adversary can instruct a corrupted player to behave arbitrarily.
An \emph{adaptive} adversary can decide which so far honest player
to corrupt at the end of a round, based on messages already sent.
However, since we will mainly consider protocols
that are secure against $n-1$ corrupted players,
adaptive corruption is not a crucial feature of the adversary.

\noindent \textbf{Ideal Cryptographical or Hardware Assumptions.}
Under the following scenarios,
we can restrict our attention to
fail-stop adversaries.
\begin{itemize}

\item \emph{Ideal Cryptographical Assumption.}
We consider adversaries that cannot break cryptographical primitives such as 
commitment schemes~\cite{DBLP:journals/jacm/LinP15} and zero-knowledge proofs~\cite{DBLP:conf/stoc/Pass04}.  At the beginning of the protocol, each player generates 
all the randomness that will be used in each round of the protocol
using \emph{verifiable random functions}~\cite{DBLP:conf/focs/MicaliRV99} and broadcast the commitments of the 
randomness, together with the corresponding zero-knowledge proofs
that the randomness and commitments are generated correctly.
Then, in each round of the protocol, 
a player uses committed randomness to generate and broadcast
the message, together with the zero-knowledge proof that
the message is generated using the committed randomness.

We remark that for Byzantine adversaries in Theorem~\ref{th:main_uniform}, 
we assume only the existence of ideal commitment schemes
(but not necessarily zero-knowledge proofs or verifiable random functions).

\item \emph{Ideal Hardware Assumption.} Each player
is assumed to reside within an SGX enclave that cannot be corrupted.
Hence, an adversary can only disrupt the broadcast channel
of a player.

\end{itemize}

We say that a protocol realizes a mechanism
if, under honest execution by all players, the protocol produces the same 
distribution of assignments 
as the mechanism.

\noindent \textbf{Player Satisfaction.}  Recall
that a mechanism returns some (random) $A \in \mathcal{A}$,
where each $A_i$, the $i$-th row of $A$, is a random vector
and the $j$-th element of expectation $\E[A_i]$ illustrates
the probability that player~$i$ gets item $j$ under this (random) mechanism.
We could easily see that $\E[A_i] \in \Delta(\mcal{M})$,
where $\Delta(\mcal{M})$ is the collection of distributions of items.
Rather than using a utility function that could give a total order for comparison,
we say the satisfaction of player~$i$
to be $\E[A_i]$
which introduces a partial order under what we call vector dominance.

\begin{definition}[Vector Dominance]
\label{defn:v_dom}
Given vectors $p, q \in \Delta(\mcal{M})$,
a player~$i$ with favorite function $O_i$ prefers $p$ to $q$ if

\begin{equation} \label{eq:v_dom}
 \forall j \in [n], \sum_{k \in [j]} p(O_i(k)) \geq \sum_{k \in [j]} q(O_i(k)).
\end{equation}

In this case, we say that $p$ dominates $q$ (with respect to~$i$),
and this defines a partial order
$p \succeq_i q$ on~$\Delta(\mcal{M})$.
Observe that the partial order can
be extended to $[0,1]^\mcal{M}$ (where
the coordinates of a vector do not necessarily sum up to 1) also via (\ref{eq:v_dom}).

\end{definition}

\begin{definition}[Matrix Dominance]
Given $P, Q \in \coA$ and $\sigma \in \mcal{S}$,
we say that $P$ dominates $Q$ (with respect to
$\sigma$), if, for all $i \in \mcal{N}$, the rows
of $P$ and $Q$ corresponding to~$i$ satisfy
$P_i \succeq_i Q_i$;
we denote this by $P \succeq_\sigma Q$.
\end{definition}

\noindent \emph{Strict Dominance.}  Observe that
we use the term ``dominate'' to refer to a binary relation $\succeq$ that happens to be reflexive; hence,
every element dominates itself.
When we say $p$ strictly dominates $q$,
we mean $p \succeq q$ and $p \neq q$.

\subsection{Some Well-Known Properties of Mechanisms}

The following property intuitively expresses the 
idea that a mechanism should return an assignment according
to the preferences of the players.

\begin{definition}[Stability]
An assignment $P \in \mcal{A}$ is stable with respect
to a preference profile $\sigma \in \mcal{S}^{\mcal{N}}$,
if there does not exist a different assignment $P'$
such that $P' \succeq_\sigma P$.

A mechanism is stable if it always produces
a stable assignment with respect to the input preference profile.
\end{definition}

\begin{definition}[Ordinal Efficiency]
A bistochastic matrix $P \in \coA$
is ordinally efficient with respect to 
$\sigma \in \mcal{S}^{\mcal{N}}$,
if there does not exist a different $P' \in \coA$
such that $P' \succeq_\sigma P$.

A mechanism is ordinally efficient
if for all inputs $\sigma \in \mcal{S}^{\mcal{N}}$,
it returns a distribution in $\Delta(\mcal{A})$
whose induced bistochastic matrix is ordinally efficient
with respect to $\sigma$.
\end{definition}

Ordinal efficiency is a stronger property than stability.
However, if a mechanism returns an assignment based on
players' preferences, then a player may benefit by lying
about its true preference.

\begin{definition}[Truthfulness]
A mechanism is (strongly) truthful,
if a player lying about its preference order 
will receive a vector in $\Delta(\mcal{M})$ that is dominated
(with respect to its true preference) 
by the vector received had it been truthful.

A mechanism is weakly truthful,
if a lying player cannot receive
a vector in $\Delta(\mcal{M})$
that strictly dominates the
vector received had it been truthful.
\end{definition}

All the properties above can be achieved 
by a deterministic mechanism (which is realized
by a trivial protocol in which no communication other than
announcing one's preference is needed).  
For instance, in a \emph{deterministic serial dictatorship}, the players can be arbitrarily ranked
and we let a higher-ranked player choose its favorite item 
before lower-ranked players.
The following property captures fairness,
and can be achieved only with randomness.


\begin{definition}[(Strong) Equal Treatment (of Equals)]
A matrix $P \in \coA$ achieves (strong) equal treatment with respect to
$\sigma \in \mcal{S}^{\mcal{N}}$ (that defines favorite functions $O_i$'s), if 
for all $i,j\in \mcal{N}$ and $\ell \in [n]$, the following holds:

``$\forall k \in [\ell], O_i(k) = O_j(k)$'' implies that
``$\forall k \in [\ell], P(i, O_i(k)) = P(j, O_j(k))$''.

A matrix $P$ achieves weak equal treatment if the above
condition holds for $\ell = n$ (but not necessarily for other values of $\ell$).
\end{definition}


\begin{fact}[Impossibility Result~\cite{DBLP:journals/jet/BogomolnaiaM01}]
\label{fact:impossible}
For $n \geq 4$ players,
there is no mechanism that can achieve all the following:
ordinal efficiency, strong truthfulness and weak equal treatment.
\end{fact}

This impossibility result implies that
any fair mechanism (in terms of equal treatment) is either (i)~not strongly truthful
or (ii)~not ordinally efficient.
In case~(i), this means that a player might have the incentive to lie about its preference order.
In case~(ii), this means that potentially all players might collude and deviate from the protocol such that everyone is better off.
Therefore, in this paper, we focus on notions that provide some guarantees to honest and truthful behavior, as opposed to discouraging deceitful or corrupted behavior.

\subsection{Security Notions of Protocols}

We introduce our security notions for protocols and explain
the intuition.
The next security notion 
encourages a player to remain honest even when there are corrupted players,
because it captures the guarantee that an honest player will not be hurt.

\begin{definition}[(Approximate) Maximin Security]
\label{defn:maximin}
For $\epsilon \geq 0$,
a protocol $\Pi$ is $(1-\epsilon)$-maximin secure
against an adversary $\Adv$ if the following holds
for any input preference profile $\sigma \in 
\mcal{S}^{\mcal{N}}$.
Given $\sigma$,
suppose that $Q \in \mcal{A}$ is the (random) assignment
produced by $\Pi$ under the strategy of $\Adv$,
while $P \in \mcal{A}$ is the one produced had every player behaved honestly.
Then, for every honest player~$i$,
the expectations of the $i$-th rows satisfy
$\E[Q_i] \succeq_i (1 - \epsilon) \cdot \E[P_i]$,
where the partial order $\succeq_i$ is 
defined in Definition~\ref{defn:v_dom} with respect to the
 preference $\sigma_i$ of player~$i$.

In this work, we focus on the special case $\epsilon = 0$, which is simply known as maximin secure.
\end{definition}

As mentioned in the introduction,
if players can lie about their preference orders,
then the best guarantee that one can only hope for
is that an honest player still
receives something that is at least as good as a uniformly random item.

\begin{definition}[Uniform Dominance]
\label{defn:u_dom}
A protocol $\Pi$ achieves uniform dominance against
an adversary \Adv if for any input
preference profile $\sigma \in \mcal{S}^{\mcal{N}}$
and any honest player~$i$, the $i$-th row
of the (random) assignment $P \in \mcal{A}$ returned 
by the protocol (under the strategy of \Adv)
satisfies
$\E[P_i] \succeq_i \mathfrak{e}$,
where $\mathfrak{e} \in \Delta(\mcal{M})$
is the uniform vector and
$\succeq_i$ is the partial order
defined in Definition~\ref{defn:v_dom} with respect to the
 preference $\sigma_i$ of player~$i$.

\end{definition}

\begin{remark}
An equivalent formulation of Definition~\ref{defn:u_dom} is that for all $\ell \in [n]$,
the probability that a truthful and honest player will receive
an item from its top $\ell$ choices is at least $\frac{\ell}{n}$,
no matter what the other players say or do.
\end{remark}

\section{Standard Building Blocks}
\label{section:blocks}

We give descriptions for some well-known primitives.
Since they are all standard, we just highlight some
important properties and give the relevant references.

\noindent \textbf{Commitment Scheme.}
Assuming the existence of one-way functions/permutations,
there is a constant-round publicly verifiable commitment
scheme~\cite{DBLP:journals/jacm/LinP15} that is perfectly correct, perfectly binding, and concurrent non-malleable.
For the purpose of understanding this paper, the reader just needs to know that
the \emph{commit phase} of the scheme allows a player to 
construct a \emph{commitment} $C$ of some secret message~$m$.
In a real-world scheme, the commitment is \emph{computationally hiding},
which means a polynomial-time adversary cannot
learn anything about the secret message from $C$.
However, for ease of exposition, we will assume that the commitment
is ideally secure and the event that the adversary
can gain extra information from the commitment has zero probability.
In the \emph{open phase}, the player can choose to open the commitment
to reveal the secret message $m$, where perfectly binding means
that it is impossible to open the commitment to any other message different from~$m$.

\ignore{
\noindent \textbf{Multi-Player Coin-Flipping Problem.}  The goal 
is for a set of $n \geq 2$ players to participate in
a (randomized) protocol that output an unbiased bit in $\{0,1\}$.
However, Cleve~\cite{} showed that even with standard cryptographical tools
(such as the aforementioned perfect commitment scheme),
for any coin fipping protocol,
 there is
a polynomial-time fail-stop adversary that controls at least $\frac{n}{2}$ players
and can create a non-zero bias in the output bit.
However, with more than $\frac{n}{2}$ honest players,
standard multi-party computation techniques~\cite{} can solve
the coin-flipping problem with zero bias, in which
case any mechanism can be realized by a protocol that is (perfectly)
maximin secure.  In this paper, we also consider the case
that the adversary can control up to $n-1$ corrupted players.
}

\noindent \textbf{Lottery Problem.} There is a set $\mcal{N}$ of $n$~players,
and the input is a probability vector $p \in \Delta(\mcal{N})$.
The goal is for the players to participate in a protocol
that determines a winner such that for each $i \in \mcal{N}$,
 player~$i$ wins with a probability~$p_i$.

\noindent \emph{Duel Protocol.}  The special case $n=2$ for rational input
probability vector can be solved by an extension of the Blum's protocol~\cite{DBLP:journals/tocs/Blum83}
that uses a commitment scheme.
On a high level, in the first round, each of the two players picks a random element
from some appropriate ring and broadcasts its commitment.
In the second round, each player opens its commitment and the sum of the opened elements determines the winner.  If a player fails to open its commitment, the other player is the winner; if both players fail to open their commitments, a default player can be the winner.
It is straightforward that an honest player~$i$ wins with a probability at least $p_i$
even if the other player is controlled by a Byzantine adversary.
However, as the duel protocol is used as a subroutine later,
there is some subtlety when both players are controlled by the adversary.
Observe that a Byzantine adversary can choose which player to be the winner
without being detected, while any deviation by a fail-stop adversary
will be immediately revealed in the transcript.  This distinction is important
later as we consider maximin security of protocols.

\noindent \emph{Tournament Tree Protocol.}  The duel protocol can be
generalized to any $n \geq 2$ players with rational input probability
vector by the \emph{tournament tree} protocol that has a binary tree
structure in which each internal node corresponds to an instance
of a duel protocol; for a detailed description, see~\cite{DBLP:conf/crypto/ChungCWS21}.
Again, any honest player~$i$ wins with a probability at least $p_i$
even if all other players are controlled by a Byzantine adversary.
Similarly, as in the duel protocol,
if all players are corrupted, a Byzantine adversary
can choose any player to be the winner without being detected.

\subsection{Augmented Protocols for the Lottery Problem}

\ignore{As remarked earlier, in the standard duel or tournament tree protocols,
if only a subset of players participate
in some instance of the protocols,
a Byzantine adversary controlling all participants of that instance
can choose the winner arbitrarily without being detected.
Hence, if we use the standard protocols as subroutines,
a Byzantine adversary might be able to change the distribution
of some outcomes without being detected, and hence, we might
be able to defend only against fail-stop adversaries.
On the other hand, if there are less than
$\frac{n}{2}$ corrupted players,
cryptographical tools~\cite{} allow us to solve coin-flipping with zero bias.}


\ignore{
When there is a prior knowledge on an upper bound
on the fraction of corrupted players,
we augment the duel protocol such that
to decide a winner
among two players, every player in~$\mcal{N}$ also participates.
In addition to the original properties, the augmented protocol guarantees
that provided that there are more than $\frac{n}{2}$ honest players,
any (Byzantine) adversary can change the output distribution only via abortion, which will be reflected in the transcript.
}

For completeness,
we describe the augmented duel protocol and introduce
the terminology to describe the detection of corrupted players,
in the case of fail-stop adversaries.

In an instance $\mathsf{AugDuel}(p_1, p_2)$,
there are non-negative integers $k_1, k_2 \in \Z$
such that for $i \in \{1, 2\}$,
player~$i$ is supposed to win
with probability $p_i = \frac{k_i}{k_1 + k_2}$.
In addition to the two players, all players in~$\mcal{N}$ (might)
participate as follows.

\begin{enumerate}

\item  \emph{Commit Step.}
Denote $k := k_1 + k_2$ and $\ell := \ceil{\log_2 k}$.
Each player~$i \in \{1, 2\}$ samples a uniformly random element $s_i$ in the
ring $\Z_k$, which can be represented by
an $\ell$-bit string; each player in $\{1, 2\}$ commits to its string and broadcasts
the commitment.

\ignore{
\item \emph{Augmented Step.} 
If there is no prior guarantee on the fraction of corrupted players,
set $\widehat{s} := 0$; otherwise,
all players in~$\mcal{N}$
use (multiple instances of) a multi-party coin
flipping protocol to sample a random $\widehat{s} \in \Z_k$.
If $\mcal{N}$ contains more than $\frac{n}{2}$ honest players,
then $\widehat{s}$ is uniformly random.
}

\item \emph{Open Step.}  Each player~$i \in \{1, 2\}$ opens
its commitment to reveal $s_i$.
If $s_1 + s_2 \in \{0, 1, \ldots, k_1 -1 \}$,
then player~1 wins; else, player~2 wins.

\item \emph{Corruption Detection and Survivor.}  If a player in~$\{1, 2\}$
aborts or fails to open its commitment to reveal
an element in $\Z_k$, then the protocol identifies
this player as \emph{corrupted}.
If there is only one identified corrupted player,
the other player is the winner;
if both players are identified as corrupted,
a default winner (say player 1) can be chosen.

A player in $\{1, 2\}$ that neither wins nor is identified as corrupted is known
as a \emph{survivor}.
\end{enumerate}

\begin{lemma}[Augmented Duel Protocol]
\label{lemma:aug_duel}
In an instance $\mathsf{AugDuel}(p_1, p_2)$
of the augmented duel protocol,
the following properties hold.

\begin{enumerate}
\item Even when \Adv is Byzantine,  an honest player~$i \in \{1, 2\}$ wins with a probability
at least $p_i$.


\item 
Suppose \Adv is fail-stop.
Then,
there exists a coupling\footnote{In probability theory,
a coupling between two probability spaces
$(\Omega_1, \Pr_1)$ and
$(\Omega_2, \Pr_2)$
is a joint space $(\Omega_1 \times \Omega_2, \Pr)$,
whose projections into $\Omega_1$ and $\Omega_2$
equal to $(\Omega_1, \Pr_1)$ and 
$(\Omega_2, \Pr_2)$, respectively.
} 
between 
the honest execution and the corrupted execution under the strategy
of \Adv
such that
if the survivor sets $S$ and
$S^{\Adv}$ correspond to the honest and the corrupted executions, respectively,
it holds that $S^{\Adv} \subseteq S$;
moreover, if an honest player wins in the honest execution,
it also wins in the corrupted execution.

\ignore{
If a (Byzantine) adversary changes
the winner distribution to $p' := (p_1', p_2')$,
then the protocol detects
corruption with probability at least $\frac{1}{2} \|p - p'\|_1$.
}
\end{enumerate}
\end{lemma}

\begin{proof}
    The first statement for Byzantine adversaries is a well-known result, and we prove the second statement under
    fail-stop adversaries.

    We first describe the coupling. We sample $s_1$ and $s_2$ independently from $\Z_k$
    and use them to create a coupling between 
    an honest execution and an execution under
    the strategy of \Adv.
    Observe that $s_0 = s_1 + s_2$ is distributed uniformly at random in $\Z_k$.

    Recall that the goal is to show that
    by fixing $s_1$ and $s_2$,
    we always have $S^{\Adv} \subseteq S$,
    where $S$ is the survivor set under honest execution.
    
    Finally, without loss of generality,  
    assume that
    conditioning on some value $s_0 = s_1 + s_2$,
    the survivor set is $S = \{1\}$, which means player~2 is the winner under honest execution.
    The only way to contradict $S^{\Adv} \subseteq S$
    is to make $2 \in S^{\Adv}$, i.e., $2$ cannot be a winner 
    under the strategy of $\Adv$.
    
    Conditioning on this value of $s_0$, observe
    that the only way the adversary $\Adv$ 
    can make player~2 lose the duel is to make it fail
    to open its commitment, thereby identifying player~2 as corrupted;
    this also means that player~2 cannot lose if it remains honest.
    
    Therefore, it follows that $2 \notin S^{\Adv}$,
    which means that $S^{\Adv} \subseteq S$;
    moreover, if $2$ is honest, then it also wins in
    the corrupted execution.
  \end{proof}

\noindent \emph{Extension to the Tournament Tree Protocol.}
We can use the augmented $\mathsf{AugDuel}$ as a subroutine
in the tournament tree protocol.
Given a subset $\mcal{N}' \subseteq \mcal{N}$,
we denote an instance of the augmented tournament tree protocol by
$\mathsf{AugTourn}(p_i: i \in \mcal{N}')$,
where each instance of the duel protocol
is implemented by $\mathsf{AugDuel}$.
Similarly, a non-winning player of $\mathsf{AugTournament}$
that
is not identified as corrupted in any 
$\mathsf{AugDuel}$ instance is known as a survivor.
A similar result is given as follows.

\begin{lemma}[Augmented Tournament Tree Protocol]
\label{lemma:aug_tree}
In an instance $\mathsf{AugDuel}(p_i: i \in \mcal{N}')$
of the augmented tournament tree protocol,
the following holds.
\begin{enumerate}
\item Even when \Adv is Byzantine, an honest player~$i \in \mcal{N}'$ wins with a probability
at least $p_i$ (even when all other players
in $\mcal{N}$ are corrupted).

\item 
Suppose that \Adv is fail-stop.
Then, there exists a coupling between 
an honest execution and the execution under the strategy
of \Adv with corresponding survivor
sets $S$ and $S^{\Adv}$ such that
it holds that $S^{\Adv} \subseteq S$;
moreover, under this coupling,
if an honest player wins under the
honest execution,
it also wins in the corrupted execution.

\end{enumerate}
\end{lemma}

\begin{proof}
    The proof follows from Lemma~\ref{lemma:aug_duel}, which gives the first
    statement.
    
    For the second statement, we apply the same coupling over all instances
    of $\mathsf{AugDuel}$ as in the proof of Lemma~\ref{lemma:aug_duel}.  Suppose in an honest execution over candidates $\widehat{\mcal{N}}$, some player~$i_0$ is the winner of $\mathsf{AugTourn}$, which means the
    survivor set is $S = \widehat{N} \setminus \{i_0\}$.  
    This means that $i_0$ is the winner of all the $\mathsf{AugDuel}$ instances.
    Under the same conditions as in Lemma~\ref{lemma:aug_duel},
    there is no way $i_0$ can lose any of the duels without being identified as
    a corrupted player.  Therefore, $i_0 \notin S^{\Adv}$ and the result follows.
    
  \end{proof}

\section{Impossibility Result to Achieve Maximin Security}
\label{section:impossible}

The impossibility result in Fact~\ref{fact:impossible} states that
in general, no mechanism can achieve strong truthfulness, ordinal efficiency and weak
equal treatment simultaneously.
Hence, even when all players are honest, no protocol can realize such a mechanism.  Recall that we have the distinction between truthfulness (whether a player reveals its true preference) and honesty (whether
a player follows a protocol), and the notion
of maximin security in Definition~\ref{defn:maximin}
is concerned about players' honesty (as opposed to their truthfulness).
Therefore, as a first step to designing protocols, it is natural to ask whether it is possible
to have a maximin secure protocol that realizes
a mechanism that satisfies ordinal efficiency and strong equal treatment.
The goal of this section is the following impossibility result.

\begin{theorem}[Impossibility Result]
\label{th:impossible}
There exists an instance with $n=4$ players such that
any mechanism that achieves ordinal efficiency
and strong equal treatment
cannot be realized by a maximin secure protocol (which terminates
in a bounded number of rounds)
against fail-stop adversaries that control
at least $\frac{n}{2}$ players.
\end{theorem}

To get some intuition about ordinally efficient mechanisms,
we revisit a well-known ordinally efficient mechanism
that also achieves strong equal treatment.


Recall that the \PS mechanism~\cite{DBLP:journals/jet/BogomolnaiaM01}
gives a (deterministic) procedure to compute the induced bistochastic matrix $P$ from a given preference profile.
Initially, all items are \emph{unconsumed}.  
At any moment, each player can fractionally consume
its favorite remaining item at a unit rate until that item
is totally consumed.  Observe that at time~1, all items will
be totally consumed.  The entry $P(i,j)$ is the fraction
of item~$j$ consumed by player~$i$ in this process.

\begin{example}[Instance with 4 Players]
\label{ex:PS}
Consider $n=4$ players with the following preference profile $\sigma$
on $\mcal{M} := \{m_i : i \in [4]\}$,
where the $\mathsf{PS}$ mechanism produces
the bistochastic matrix $P_{\mathsf{PS}} \in \coA$.

\begin{minipage}{0.48\textwidth} 
$$m_1 \succ_1 m_3 \succ_1 m_2 \succ_1 m_4$$
$$m_1 \succ_2 m_4 \succ_2 m_2 \succ_2 m_3$$
$$m_2 \succ_3 m_3 \succ_3 m_1 \succ_3 m_4$$
$$m_2 \succ_4 m_4 \succ_4 m_1 \succ_4 m_3$$
\end{minipage}
\begin{minipage}{0.48\textwidth}
$P_{\mathsf{PS}} = \left[ \begin{matrix}
\frac{1}{2} & 0 & \frac{1}{2} & 0\\
\frac{1}{2} & 0 & 0 & \frac{1}{2} \\
0 & \frac{1}{2}  & \frac{1}{2} & 0\\
0 & \frac{1}{2}  & 0 & \frac{1}{2} 
\end{matrix} \right]$ 
\end{minipage}

\vspace{5pt}

Observe that $P_{\mathsf{PS}}$ corresponds to a
unique distribution $\Delta(\mcal{A})$ as follows:
$P_{\mathsf{PS}} = \frac{1}{2} A_{\mathsf{head}}
+ \frac{1}{2} A_{\mathsf{tail}}$, where

\begin{center}
$A_{\mathsf{head}} = \left[ \begin{matrix}
0 & 0 & 1 & 0\\
1 & 0 & 0 & 0 \\
0 & 1  & 0 & 0\\
0 & 0  & 0 & 1 
\end{matrix} \right]$
and
$A_{\mathsf{tail}} = \left[ \begin{matrix}
1 & 0 & 0 & 0\\
0 & 0 & 0 & 1 \\
0 & 0  & 1 & 0\\
0 & 1  & 0 & 0 
\end{matrix} \right]$. 
\end{center}

\end{example}

\begin{lemma}[Unique Distribution]
\label{lemma:unique}
For the problem instance in Example~\ref{ex:PS},
$P_\mathsf{PS}$ is the unique bistochastic matrix
that achieves both ordinal efficiency
and strong equal treatment.
\end{lemma}

\begin{proof}
Suppose $P \in \coA$ is a bistochastic
matrix that achieves both ordinal efficiency and
strong equal treatment for the instance in Example~\ref{ex:PS}.

First, consider player~1.
Because of strong equal treatment, 
we have $P(1,m_1) = P(2,m_1)$, which implies
that $P(1,m_1) \leq \frac{1}{2} = P_{\mathsf{PS}}(1, m_1)$.
Since $P_{\mathsf{PS}}(1, O_1(1)) + P_{\mathsf{PS}}(1, O_1(2))
= P_{\mathsf{PS}}(1, m_1) + P_{\mathsf{PS}}(1, m_3)
 = 1$,
we have for $\ell \in \{2,3,4\}$,
$1 = \sum_{j\in [\ell]} P_{\mathsf{PS}}(1, O_1(j))
\geq \sum_{j\in [\ell]} P(1, O_1(j))$.

It follows that the rows of the matrices corresponding
to player~1 satisfy:  $P_{\mathsf{PS}}(1, \cdot) \succeq_1 P(1, \cdot)$.

A similar analysis for players 2 to 4 implies that
$P_{\mathsf{PS}} \succeq_\sigma P$,
which means $P_{\mathsf{PS}}$ dominates $P$ with respect to $P$.
Since $P$ is ordinally efficient with respect to $\sigma$,
it follows that $P = P_{\mathsf{PS}}$.
  \end{proof}

\subsection{Reduction from Coin-Flipping Problem}

We next complete the proof of Theorem~\ref{th:impossible}.
We assume that there is a maximin secure protocol $\Pi$
that terminates within a bounded number of rounds, and realizes a mechanism that achieves both equal treatment
and ordinal efficiency on the instance in Example~\ref{ex:PS}.
From $\Pi$, we will construct a two-party coin-flipping protocol.
Finally, we show that Cleve's impossibility result~\cite{DBLP:conf/stoc/Cleve86}
will contradict the maximin security of $\Pi$
(against fail-stop adversaries controlling at most 2 players).

\noindent \textbf{Interpreting $\Pi$ as
a Two-Party Coin-Flipping Protocol.}  Suppose party~$A$
and part~$B$ would like to use $\Pi$ in Example~\ref{ex:PS} as a coin-flipping
protocol.  Party~$A$ controls players in $\{1,2\}$,
while party~$B$ controls players in $\{3, 4\}$.
Observe that we consider the case that at most one party is corrupted
by a fail-stop adversary.
By Lemma~\ref{lemma:unique}, when both parties are honest,
the outcome of $\Pi$ can be either $A_{\mathsf{head}}$
or $A_{\mathsf{tail}}$, which can be naturally interpreted
as a coin outcome.  However, if either party is corrupted,
there can be other outcomes of $\Pi$ that we need
to interpret as coin outcomes, after which the
description of the coin-flipping protocol will be completed.

The following lemma says that the maximin security of $\Pi$ implies that when only one party is corrupted, the assignment for players in the honest party still satisfies either assignment matrix $A_{\mathsf{head}}$ or $A_{\mathsf{tail}}$.
Hence, if possible, we can use the assignment for players in either party~$A$ or party~$B$ to determine the coin outcome.  Note that
the assignments for the two parties will not contradict each other.
The reason is that combining the first two rows from one of the assignment matrices
and the last two rows from the other assignment matrix will not be a valid 
assignment.

Finally, if the assignment for the players in neither party
is consistent with $A_{\mathsf{head}}$ or $A_{\mathsf{tail}}$,
Lemma~\ref{lemma:honest_party} implies that both parties are corrupted,
in which case any default coin outcome (say $\mathsf{head}$) can be returned.

\begin{lemma}[Assignment for an Honest Party]
\label{lemma:honest_party}
Suppose $\Pi$ is maximin secure against a fail-stop
adversary controlling at most 2 players.
Then, the assignment for players in an honest party in $\Pi$
agrees with either $A_{\mathsf{head}}$
or $A_{\mathsf{tail}}$, each of which happens with probability $\frac{1}{2}$.
\end{lemma}

\begin{proof}
We consider the case that party~$A = \{1,2\}$ is honest, and the case when party~$B$ is honest can be analyzed similarly.

By the maximin security of $\Pi$,
each player in $\{1, 2\}$ receives $m_1$ with probability $\frac{1}{2}$
(which means which player receives $m_1$ partitions
the sample space into two equally likely events).
Since the vector received by player~1 
must dominate the row $P_{\mathsf{PS}}(1, \cdot)$
(with respect to its own preference),
it follows that if player~1 does not receive $m_1$ (which happens
with probability $\frac{1}{2}$), it must receive $m_3 = O_1(2)$.

Similarly, if player~2 does not receive $m_1$,
then it must receive $m_4 = O_2(2)$.
Therefore, it follows that the assignment for
players in party~$A$ satisfies either 
$A_{\mathsf{head}}$ or $A_{\mathsf{tail}}$,
each of which happens with probability~$\frac{1}{2}$.
  \end{proof}

\begin{corollary}[Contradiction to Cleve's Result~\cite{DBLP:conf/stoc/Cleve86}]
Lemma~\ref{lemma:honest_party} implies that
when there is only one corrupted party in the two-party coin flipping protocol,
the outcome of the coin is unbiased.
\end{corollary}

\section{Achieving Perfect Maximin Security}
\label{sec:algo}

In view of the impossibility results in
Fact~\ref{fact:impossible}
and Theorem~\ref{th:impossible},
we design a protocol (assuming ideal cryptographical tools) that achieves the following
properties.

\begin{theorem}[Achieving Stability, Strong Equal Treatment and Maximin Security]
\label{th:maximin}
Assuming an ideal commitment scheme,
there exists a protocol that realizes 
a mechanism that achieves stability and strong equal
treatment (when all players behave honestly);
moreover, the protocol achieves
perfect maximin security against a fail-stop adversary
that controls up to $n-1$ players.
\end{theorem}

\subsection{Preference Priority (\PP) Protocol}

\noindent \textbf{Algorithm Intuition.}
The \PP protocol in Algorithm~\ref{alg:PP} proceeds according to round $r$
from $1$ to $n$.
In round~$r$, survivors (that have not
been assigned their at least $(r-1)$-st items) will 
compete for their $r$-th favorite items (if still available) via the $\mathsf{AugTourn}$ protocol. If the $r$-th favorite item for a player is no longer available,
then the player does not compete for any item in round~$r$.  Observe that
this implies that the protocol is not weakly truthful (in the
case where all players honestly follow the protocol).
However, this rigidity of when an item can be assigned is how
this protocol achieves maximin security.


\begin{algorithm}[!ht]
\caption{Preference Priority Protocol}
\label{alg:PP}
\textbf{Input:} A preference profile $\sigma \in \mathcal{S}^{\mathcal{N}}$ (declared by players) for the items~$\mathcal{M}$.\\
\textbf{Output:} A (random) assignment $A \in \mathcal{A}$. \\

\textbf{Initialization:} \\

Set $A$ to an empty assignment (i.e., a zero matrix). 

Let $S \gets \mathcal{N}$ denote the current collection of valid survivors.

Let $R \gets \mathcal{M}$ denote the current collection of available items.

\For{$r$ $\mathrm{from}$ $1$ $\mathrm{to}$ $n$}{

    Let $\mcal{C} = \{ \{i\in S: O_i(r)=j\}: j \in R\}$ 
		be a \emph{partial} partition of $S$ according to each survivor's $r$-th favorite item in $R$. \label{line:C}
		


    \ForEach{$C \in \mcal{C}$ in an arbitrary order}{

				Suppose for all $i \in C$, the common $r$-th favorite item is $O_i(r) = j$.
				
        Let $\vec{p}=(p_i = \frac{1}{|C|}: i\in C)$ indicate that players in $C$ should compete for item $j$ uniformly at random.\\

        Run $\textsf{AugTourn}(\vec{p})$ to obtain the winner $\widehat{i}$ and the survivor set $S'\subseteq C \setminus \{\widehat{i}\}$;
				if $|C|=1$, we assume that the only player in $C$ is the default winner
				and cannot abort.\\
       
        Assign $A(\widehat{i},j)\leftarrow 1$.\\
        
        Update  $R \gets R \setminus \{j\}$ and $S \gets S\setminus (C \setminus S')$.
        
    }
    
}

Any remaining items are arbitrarily assigned to players with no items yet (according to some pre-determined rule) and update A accordingly.

\textbf{return} assignment $A$
\end{algorithm}

\noindent \textbf{Naive Variant.}
Observe that one could consider a more straightforward
variant of 
Algorithm~\ref{alg:PP}.
In each round $r$, instead of restricting the survivors to compete for their $r$-th favorite item, we allow them to compete for their most preferred remaining items.
Specifically, we replace $\mathcal{C}$ in line~\ref{line:C} in Algorithm~\ref{alg:PP} by partitioning $S$ according to each survivor's favorite item in~$R$,
i.e., we denote
$O_i(R)$ as the favorite item of $i$ in $R$,
and let $\mcal{C} = \{ \{i\in S: O_i(R)=j\}: j \in R\}$.
We call this variant of the algorithm $\mathsf{NaivePP}$.

One can verify that the following Lemma~\ref{lemma:pp-strong-equal-and-stable} is still valid for $\mathsf{NaivePP}$. 

\begin{lemma}\label{lemma:pp-strong-equal-and-stable}
The \PP protocol described in Algorithm~\ref{alg:PP}
realizes a mechanism that achieves strong equal treatment
and stability.
\end{lemma}

\begin{proof}
Strong equal treatment follows because
if two players have exactly the same preference
for their $r$ most favorite items,
then they will behave in exactly the same way as
long as those $r$ items are not assigned.

Stability follows because if a player is assigned an item $j$ in the $r$-th round,
then all its more preferred items than~$j$ are no longer available
at the beginning of the $r$-th round.
  \end{proof}

However,
the following Example~\ref{eg:naive-pp-not-maximin} shows that
$\mathsf{NaivePP}$ is not maximin secure.

\begin{example}[$\mathsf{NaivePP}$ Not Maximin Secure]\label{eg:naive-pp-not-maximin}
  Consider
  $n=9$ players with
   the preference profile given by the following matrix,
  where each row corresponds to a player and each entry~$(i,j)$
  contains the index of the $j$-th favorite item for player~$i$.

  \[\left[
  \begin{matrix}
  1 & 2 & 3 & 4 & 5 & 6 & 7 & 8 & 9 \\
  1 & 2 & 6 & 4 & 5 & 7 & 8 & 9 & 3 \\
  1 & 2 & 6 & 4 & 5 & 7 & 8 & 9 & 3 \\
  1 & 2 & 6 & 4 & 5 & 7 & 8 & 9 & 3 \\
  1 & 2 & 6 & 4 & 5 & 7 & 8 & 9 & 3 \\
  5 & 2 & 3 & 7 & 8 & 6 & 9 & 1 & 4 \\
  5 & 2 & 3 & 7 & 8 & 6 & 9 & 1 & 4 \\
  5 & 3 & 6 & 7 & 8 & 9 & 1 & 2 & 4 \\
  5 & 3 & 6 & 7 & 8 & 9 & 1 & 2 & 4 \\
  \end{matrix}
  \right]\]
  
  \noindent \textit{Honest scenario.}
  We first argue that when every player
  participates honestly, player~1 will surely
  receive one of its top~4 favorite items.
  The reason is that player~1 will definitely not get item~$m_3$, since $m_3$ will be assigned to one of player~8 or player~9 at round~2. 
  This means if player~1 has not received its top~2 items by the beginning of round~3,
  it will be the only player to compete for~$m_4$ in round~3.
  
  \noindent \textit{Corrupted scenario.}
  We show that if both players~8 and player~9 abort in their first round,
  then with positive probability, player~1 does not
  get one of its top~4 favorite items.
  
  Observe that with positive probability, player $7$ gets $m_5$. Then, with positive probability both player $1$ and player $6$ fail to get $m_2$, and hence these two players will compete for $m_3$ at round $3$. With positive probability, player $1$ fails to receive its top 3 favorite items, and it has to compete with at least one player (from players $2$ to $5$) for $m_4$ at round $4$. Hence, we conclude that when player $8$ and player $9$ abort in their first round, the probability of player $1$ getting one of its top 4 favorite items is less than $1$.
  \end{example}

\subsection{Maximin Security Analysis}

\noindent \textbf{Simplifying Notation.} We first introduce some notations
to facilitate the analysis of maximin security.
Throughout the analysis,
we fix some input preference profile $\sigma$ and honest player~$i_0 \in \mcal{N}$,
and use $\succeq$ to denote the partial
order on $\Delta(\mcal{M})$ defined
in Definition~\ref{defn:v_dom} with respect to the
preference $\sigma_{i_0}$ of player~$i_0$.

\noindent \textbf{Execution State.} We use $\Lambda$ to denote
the collection of execution states of the protocol in Algorithm~\ref{alg:PP}.
A state $\lambda = (r, S, R, a) \in \Lambda$ 
is a tuple, where the protocol is currently at the 
beginning of round $r \in [n]$,
$S$ is the current collection of survivors,
$R$ is the current collection of remaining items,
and if $i_0 \notin S$, then $a \in \mcal{M}$ is the item already received by $i_0$,
and $a = \bot$ otherwise.

While we consider a state~$\lambda$ that may not be reachable from an honest execution,
we only consider \emph{valid} states that satisfy the following conditions:
\begin{itemize}
\item $|S| \leq n - r + 1$ and $|S| \leq |R|$;
\item if $i_0 \in S$, then all of the $(r-1)$ most favorite items
of $i_0$ are not in $R$.
\end{itemize}

Observe that if $i_0$ is honest,
then only valid states (which are defined with respect to $i_0$) can be reached in the execution of the protocol.


\noindent \textbf{Item Distribution.}
For any state $\lambda \in \Lambda$,
we use $\Pi(\lambda) \in \Delta(\mcal{M})$
to denote the distribution of the item received by player~$i_0$
if the protocol is executed honestly by all players from state~$\lambda$ onwards.
Since $\Pi(\lambda)$ has no randomness if $i_0 \notin S$ (because
$i_0$ has already received an item from $A$),
it suffices to consider the case $i_0 \in S$ (and $a = \bot$).

\begin{lemma}[Monotonicity with Respect to Removing Survivors or Adding Items]
\label{lemma:mono_survivors}
Consider a valid state 
$\lambda = (r, S, R, \bot) \in \Lambda$,
where $S$ contains an honest player $i_0$.
Then, the following monotone properties hold.

\begin{compactitem}

\item[$P(r)$:] \emph{Remove Player.}
Suppose $i \neq i_0$ is another player,
and $\lambda_1 = (r, S \setminus \{i\}, R, \bot)$, 
where we allow $i \notin S$.
Then, $\Pi(\lambda_1) \succeq \Pi(\lambda)$.

\item[$Q(r)$:] \emph{Add Item.}
Suppose $j \in \mcal{M} \setminus R$ is an item not in $R$
such that $\lambda_2 = (r, S, R \cup \{j\}, \bot)$ is still valid.
Then, $\Pi(\lambda_2) \succeq \Pi(\lambda)$.

\end{compactitem}

\end{lemma}

\begin{proof}
  We consider backward induction on~$r$.
  For the base case $r = n$,
  because of the definition of a valid state,
   we have the 
  trivial case that $S$ contains a single survivor $i_0$
  and $R$ contains a single item $O_{i_0}(n)$.
  Hence, $\lambda = \lambda_1$, and the
  statement $P(n)$ holds.  Observe that there is no
  other item $j \notin R$ such 
  that $\lambda_2 = (n, S, R \cup \{j\}, \bot)$ is also valid;
  hence, the statement $Q(n)$ also trivially holds.
  
  Consider some $1 \leq r < n$ such that
  for all $r+1 \leq t \leq n$, the statements
  $P(t)$ and $Q(t)$ are true.
  
  We first prove the statement $P(r)$.  Suppose player~$i \neq i_0$ is removed.
  Since the case $i \notin S$ is trivial, it suffices to consider $i \in S$.
  Suppose $j = O_i(r)$ is the $r$-th favorite item of player~$i$.
  If $j \notin R$, then player~$i$ is not going to compete
  for any item in round~$r$, and so this is equivalent to 
  removing player~$i$ in round~$r+1$, and we can use $P(r+1)$;
  hence, we can assume $j \in R$.
  
  Suppose $C = \{s \in S: O_s(r) = j\}$ are survivors
  that compete for item~$j$ in this round~$r$.
  Observe that survivors in $S$ not competing for~$j$ behave
  the same in round~$r$ in states $\lambda = (r, S, R, \bot)$ and 
  $\lambda_1 = (r, S \setminus \{i\}, R, \bot)$.
  There are two sub-cases.
  
  \begin{itemize}
  
  \item $|C| = 1$. This means in the current round~$r$,
  no other survivor in $S$ views $j$ as its $r$-th favorite item.
  In particular, this implies that item~$j$ is not
  within the $r$ most favorite items of player~$i_0$.
  We can construct a coupling between the states at
  the beginning of round~$r+1$ resulting from $\lambda$
  and $\lambda_1$.  For every state $\widehat{\lambda} = (r+1, \widehat{S}, \widehat{R}, a)$
  that results from $\lambda$,
  we map it to $\widehat{\lambda}_1 = (r+1, \widehat{S}, \widehat{R} \cup \{j\}, a)$
  that results from $\lambda_1$;
  observe that both transitions occur with the same probability.
  Hence, the statement $Q(r+1)$ implies that $P(r)$ is true in this case.
  
  \item $|C| \geq 2$.  We construct a coupling
  between states resulting from $\lambda$ and $\lambda_1$.
  Suppose from $\lambda$,
  the next state at the beginning of round~$r+1$
  is $\widehat{\lambda} = (r+1, \widehat{S}, \widehat{R}, a)$.
  There are two further cases.
  
  \noindent \emph{(i)} Case $i \in \widehat{S}$. 
  This means player~$i$ did not win
  in $\mathsf{AugTourn}$ for item~$j$.
  In this case, we map $\widehat{\lambda}$
  to $\widehat{\lambda}_1 = (r+1, \widehat{S} \setminus \{i\}, \widehat{R}, a)$,
  and use $P(r+1)$ in this case.
  
  \noindent \emph{(ii)} Case $i \notin \widehat{S}$.
  This means player $i$ has won item~$j$ starting from state~$\lambda$.
  To create the coupling under this case,
  from state~$\widehat{\lambda}_1$, we pick a player~$i' \in C \setminus \{i\}$ uniformly 
  at random to be the winner of item~$j$.
  The result is the state $\widehat{\lambda}_1 = (r+1, \widehat{S} \setminus \{i'\},
  \widehat{R}, a)$, and we also use $P(r+1)$ for this case.
  
  \end{itemize}
  
  We next prove the statement $Q(r)$, i.e.,
  we add item $j \notin R$ at the beginning of round~$r$.
  Let $r^*$ be the minimum round at least $r$
  such that there exists $i \in S$ such that $O_i(r^*) = j$.
  If $r^* > r$, then we can use the statement $Q(r^*)$;
  hence, we can assume that there is some $i \in S$ such that
  $O_i(r)=j$,
  and so $C = \{i \in S: O_i(r) = j\}$ is non-empty.
  
  We create a coupling between states at the 
  beginning of round $r+1$ resulting
  from $\lambda = (r, S, R, \bot)$
  and $\lambda_2 = (r, S, R \cup \{j\}, \bot)$, respectively.
  
  Observe that if $\widehat{\lambda} = (r+1, \widehat{S}, \widehat{R}, a)$
  results from $\lambda$,
  then $C \subseteq \widehat{S}$ because players in $C$ did not
  compete for any item in round~$r$.
  However, from state $\lambda_2$,
  exactly one of $C$ will win item~$j$ and be removed from $S$;
  hence, we randomly pick one player $i \in C$ to win item~$j$.
  There are two cases.
  
  \noindent \emph{(i)} Case $i = i_0$. Observe
  that player~$i_0$ prefers item~$j$ to any item in $R$.
  Hence, in this case $\widehat{\lambda}_2 = (r+1, \widehat{S} \setminus \{i_0\}, 
  \widehat{R}, j)$ is better for $i_0$ than $\widehat{\lambda}$.
  
  \noindent \emph{(ii)} Case $i \neq i_0$.
  In this case, we consider $\widehat{\lambda}_2 = (r+1, \widehat{S} \setminus \{i\},
  \widehat{R}, a)$, and apply the statement $P(r+1)$.

  This concludes the induction proof.
  \end{proof}

\begin{lemma}[Maximin Security]
The \PP protocol in Algorithm~\ref{alg:PP} is maximin secure
against a fail-stop adversary that controls up to $n-1$ players.
\end{lemma}

\begin{proof}
We show that an honest player~$i_0$
cannot be harmed by a fail-stop adversary \Adv.
Specifically, we argue that
if $v \in \Delta(\mcal{M})$ is the distribution vector
received by $i_0$ under the honest execution
and $v^{\Adv}$ is the corresponding one under the strategy of $\Adv$,
then $v^{\Adv} \succeq v$ with respect to the preference of~$i_0$.

Observe that Algorithm~\ref{alg:PP} consists
of multiple instances of $\mathsf{AugTourn}$.  There are two cases.

\begin{compactitem}
\item Suppose $i_0$ participates in an instance of $\mathsf{AugTourn}$
for some item~$j$ in some round~$r$.
Observe that at this moment, item~$j$ is the most preferred item among
the remaining items by~$i_0$.
Lemma~\ref{lemma:aug_tree} states that
the probability that $i_0$ wins item~$j$ cannot be decreased by the
adversary~$\Adv$.  If the strategy of $\Adv$
causes any corrupted player to abort in this instance of $\mathsf{AugTourn}$,
the resulting distribution received by~$i_0$ still dominates
the original distribution.

\item Suppose $i_0$ does not participate in an instance of $\mathsf{AugTourn}$.
Lemma~\ref{lemma:aug_tree} states that there exists a coupling between
an honest execution and an execution under the strategy of $\Adv$ such that
the survivor set $S^{\Adv}$ produced under $\Adv$ is
always a subset of $S$ under honest execution.  Lemma~\ref{lemma:mono_survivors}
states that removing other players cannot harm the honest player~$i_0$.
Hence, the resulting distribution received by~$i_0$ under $\Adv$ still dominates
that produced under an honest execution.
\end{compactitem}

Performing a hybrid argument on every instance of $\mathsf{AugTourn}$ in Algorithm~\ref{alg:PP} gives the required result.
  \end{proof}

\section{Achieving Uniform Dominance}
\label{sec:uniform_dom}

\noindent \textbf{Re-Visiting \PS mechanism.}
We can see from the proof of Theorem~\ref{th:impossible}
that the hurdle in realizing the $\mathsf{PS}$ mechanism
is that it can capture the coin-flipping problem,
for which a fail-stop
adversary controlling at least $\frac{n}{2}$ players
can cause a non-zero bias on the outcome probability (from the ideal~$\frac{1}{2}$).
Even though we do not know how to modify $\mathsf{PS}$
to achieve maximin security, it has inspired us
to design a variant of the protocol that 
can achieve uniform dominance against Byzantine adversaries.

\subsection{$\mathsf{OnlinePSVar}$ Protocol}

\noindent \textbf{Protocol Design Intuition.}  
The \emph{online probabilistic serial with varying rates} ($\mathsf{OnlinePSVar}$)
protocol in Algorithm~\ref{alg:onlineps} is based on the original $\mathsf{PS}$ mechanism,
but as soon as an item is totally consumed,
it is rounded via the lottery problem according to the fractional consumptions
by the players.
Observe that to make the description intuitive,
the consumption of an item seems to be an ``action'' by a player.
However, this is actually performed automatically according
to the input preference profile (about which a player could still lie though).
The players actually only actively participate in instances
of $\mathsf{AugTourn}$, which
is assumed to take zero ''time'' in the
consumption process.

We would like to highlight an important feature:
when line~\ref{ln:rate} ``Varying Rate'' is executed,
the rate of the winner in $\mathsf{AugTourn}$ is distributed among survivors proportional to their winning probabilities.
We shall see that this is extremely important to achieve
uniform dominance,
as illustrated later in Example~\ref{ex:OnlinePS}.

\noindent \emph{Terminology.} Strictly speaking, $\mathsf{OnlinePSVar}$
is a protocol.  However, when we say the $\mathsf{OnlinePSVar}$ mechanism,
we mean the corresponding mapping that takes
an input preference profile and returns a distribution of assignments when all
players behave honestly in the protocol.

\begin{algorithm}[!ht]
\caption{$\mathsf{OnlinePSVar}$ Protocol}
\label{alg:onlineps}
\textbf{Input:} A preference profile $\sigma \in \mcal{S}^{\mcal{N}}$ (declared by players) for
the items~$\mcal{M}$.\\
\textbf{Output:} A (random) assignment $A \in \mcal{A}$. \\

\textbf{Initialization:} \\

Set $A$ to an empty assignment (i.e., a zero matrix).

All items in $\mcal{M}$ are unconsumed.

\textbf{for} each player $i \in \mcal{N}$, set consumption rate of $s_i \gets 1$
unit of item per unit time.


Consider each item as 1 unit of an infinitely divisible commodity;
initialize time = 0.\\

\While{$\exists i \in \mcal{N}: s_i > 0$}
{
	Every player~$i \in \mcal{N}$ consumes
	its favorite item (according to its preference order $\sigma_i$) that is still
	not totally consumed at rate $s_i$.
	
	When an item~$j \in \mcal{M}$ is totally consumed
	by some subset $\widehat{\mcal{N}}$ of players (if there is more
	than one such item, process each item independently), do the following:
	
	Let $\vec{p} = (p_i: i \in \widehat{\mcal{N}})$ describe how players
	fractionally divide item~$j$.
	
	Run $\mathsf{AugTourn}(\vec{p})$ to obtain the winner~$\widehat{i}$
	and the survivor set $S \subseteq \widehat{N} \setminus \{\widehat{i}\}$;
	recall that a survivor is a non-winning player that is not identified as corrupted
	in $\mathsf{AugTourn}$. \label{ln:tree}
	
	Assign $A(\widehat{i},j) \gets 1$.

		%
	%
	
	\framebox{\emph{Varying Rates:}
	\textbf{for} $i \in S$, set  $s_i \gets s_i + s_{\widehat{i}} \times \frac{p_i}{\sum_{k \in \widehat{N} \setminus \{\widehat{i}\}} p_k}$ . \label{ln:rate}
	}
	
	\textbf{for} $i \in \widehat{N} \setminus S$, set rate $s_i \gets 0$.
	
		%
	%

}

Any remaining items are arbitrarily assigned to players with no items yet 
(according to some pre-determined rule) and update $A$ accordingly.

\textbf{return} assignment $A$
\end{algorithm}

\begin{lemma}[Obvious Properties]
The $\mathsf{OnlinePSVar}$ mechanism achieves  strong equal treatment
and stability.
\end{lemma}

\begin{proof}
Recall that to consider the properties of the mechanism,
we investigate what happens when all players are honest.

From the description
in Algorithm~\ref{alg:onlineps}, for two players
with exactly the same preference for their top~$k$ items,
before all those $k$ items are totally consumed or one
of them is assigned an item, 
they will behave in exactly the same way.
This implies that the mechanism achieves strong equal treatment.

Stability is also obvious because the first item that is totally consumed
will be the top choice for the player that receives it.  Applying
this observation repeatedly to the remaining items gives the conclusion.
  \end{proof}

\noindent \textbf{Varying Rates.}  
Observe that the re-distribution of consumption rates among survivors
after $\mathsf{AugTourn}$
and the detection of corrupted players
 can make the process very complicated.
Surprisingly, we have observed the following structural
property of the process.  In retrospect,
we could have replaced line~\ref{ln:rate} ``Varying Rates'' with a much
simpler updating rule, but this will make the description less intuitive.

\begin{lemma}[Consumption Rate at Joining Time]
\label{lemma:join_rate}
Consider the consumption process in $\mathsf{OnlinePSVar}$.
Suppose a player 
starts the consumption of an item at time~$t \geq 0$
(with a positive rate).
Then, we must have $t < 1$
and its consumption rate for that item is $\frac{1}{1 - t}$.
This holds even when all players are controlled by a Byzantine adversary.
\end{lemma}

\begin{proof}
	Observe that the consumption process
	can be affected by the outcome of each $\mathsf{AugTourn}$ instance
	(which has at most $n$ outcomes).  Hence,
	there are at most $n^n$ scenarios of the process,
	in which each scenario has at most $n$ possible times
	that a player can join the consumption of an item.
	Therefore, there are only a finite number of times a player
	can join the consumption of an item, and we can prove the result
	by induction on the joining time.
	
	The base case $t=0$ is trivial, because
	initially all players have a consumption rate of 1.
	For the induction hypothesis, suppose that some
	player~$i$ joins the consumption of an item~$j$
	at some time~$t > 0$ such that if any player
	starts joining the consumption of any item at time $t' < t$,
	it must be the case that $t' < 1$ and the rate
	at that moment is $\frac{1}{1 - t'}$.
	
	Since $t > 0$, this means
	that player~$i$ has just finished
	participating in some $\mathsf{AugTourn}$
	for another item~$\widehat{j}$, and is a (non-winning) survivor.
	Consider the winner~$d$ for item~$\widehat{j}$,
	and let $t_d$ and $t_i$ be the corresponding
	joining times.
	
	By the induction hypothesis, both $t_d$ and $t_i$ are less than 1, and
	the two players are consuming item $\widehat{j}$
	at rates $s_d = \frac{1}{1 - t_d}$
	and $s_i = \frac{1}{1 - t_i}$.
	This means that in the instance of $\mathsf{AugTourn}$
	for item~$\widehat{j}$,
	their winning probabilities
	are $p_d = (t - t_d) s_d$ and
	$p_i = (t - t_i) s_i$.
	
	Therefore, the rate re-distribution rule 
	gives that after $\mathsf{AugTourn}$,
	the new rate for player~$i$ is:
	
	$s = s_i + s_d \cdot \frac{p_i}{1 - p_d} = s_i + \frac{s_d \cdot (t - t_i) s_i}{1 - (t - t_d) s_d} = s_i \cdot \frac{1 - (t - t_d) s_d + s_d \cdot (t - t_i) }{1 - (t - t_d) s_d} = \frac{1}{1 - t}.$
	
	Since $s > 0$, we must have $t < 1$ and the inductive step is completed.
	Observe that no assumption about truthfulness or honesty is needed in this proof.
  \end{proof}

\begin{corollary}[Same New Rate for Survivors of $\mathsf{AugTourn}$]
\label{cor:rates}
In $\mathsf{OnlinePSVar}$,
any instance of $\mathsf{AugTourn}$ with more than one candidate
is completed strictly before time 1, and all the resulting
survivors have the same new consumption rates.
This holds even when all players are controlled by a Byzantine adversary.
\end{corollary}

\subsection{$\mathsf{OnlinePSVar}$ Achieves Uniform Dominance}

We first illustrate that the variant
$\mathsf{OnlinePS}$ (which
is the variant of Algorithm~\ref{alg:onlineps}
with line~\ref{ln:rate} ``Varying Rates'' removed)
 does not achieve
uniform dominance (even when all players are honest).

\begin{example}[$\mathsf{OnlinePS}$ does not achieve uniform dominance]
\label{ex:OnlinePS}
Consider $n = 5$ with $A = \{1, 2, 3\}$
and $B = \{4, 5\}$, where $m_1$ and $m_2$
are the top two items for all players,
but $m_1 \succ_A m_2$ and $m_2 \succ_B m_1$.
Then, a simple calculation shows
that the probability that player~1 receives item $m_1$ or $m_2$
is: $\frac{1}{3}+\frac{2}{3}\times \frac{1}{12} < \frac{2}{5}$.
\end{example}

\noindent \textbf{Interrupted Process and Claiming Ownership in $\mathsf{OnlinePSVar}$.}  Later in our proofs,
we would like to consider the probability of whether an honest player~$i_0$ has already been
assigned an item by some time $t \in [0,1]$.  However,
it is possible that at time $t$, a player is still consuming some item~$j$.  We introduce
the concept of \emph{claiming ownership}.
The interpretation here is that we interrupt the process at this time,
and sample a Bernoulli random variable with the parameter equal
to the fraction of item~$j$ already consumed by player~$i$
to determine whether player~$i_0$ should receive that item.
Hence, when we say that
a player has \emph{claimed} the ownership of an item by time~$t$,
we mean that the player has either received that item or
has a claim to that item via the above Bernoulli process.
Observe that for $t=1$, the claim of ownership and the actual assignment are equivalent.

\begin{lemma}[Probability of Ownership]
\label{lemma:ownership}
Suppose an honest player~$i_0$ is about
to start consuming some item at time~$t_0$ (which
means at this point~$i_0$ still has not received any item).
Then, for all $t \in [t_0, 1]$,
the probability that $i_0$ will have claimed ownership
of any item by time~$t$ is $\frac{t - t_0}{1 - t_0}$.
\end{lemma}

\begin{proof}
As argued in Lemma~\ref{lemma:join_rate},
the set of possible times that some player starts
to consume some item is finite.  Hence,
we prove the result by (backward) induction starting
from larger values of~$t_0$.

Suppose~$t_0$ is the largest time
that any player can start to consume an item,
and player~$i_0$ happens to start consuming an item at time~$t_0$.
By Lemma~\ref{lemma:join_rate},
player~$i_0$ is consuming that item at rate $\frac{1}{1 - t_0}$.
The maximality of $t_0$ means that this is the last ever item available
to~$i_0$
and there cannot be another player competing with~$i_0$;
otherwise, it would have been possible to start consuming another item
at a time later than $t_0$.
Hence, by time~$t \in [t_0, 1]$,
player~$i_0$ would have consumed $\frac{t - t_0}{1 - t_0}$
fraction of item~$j$, which is exactly
the probability of claiming ownership by the Bernoulli process
at this moment.

For the induction hypothesis, suppose an honest player~$i_0$ starts to
consume some item~$j$ at some time $t_1$ such that
for all $t_0 > t_1$, the required result holds if an honest
player starts to consume an item at time~$t_0$.

By Lemma~\ref{lemma:join_rate},
$i_0$ consumes item~$j$ at rate $\frac{1}{1 - t_1}$.
Observe that if item~$j$ is not fully consumed by time $t$,
then the same argument as the base case holds,
and the probability that $i_0$ claims ownership
of $j$ by time~$t$ is $\frac{t-t_1}{1 - t_1}$.

Otherwise, we let the consumption process
carry on until item~$j$ is fully consumed at some time $t_0 > t_1$.
By Lemma~\ref{lemma:join_rate},
at time $t_0$, the fraction of item~$j$ consumed by~$i_0$ is $\frac{t_0 - t_1}{1 - t_1}$,
which is the probability that (honest) player $i_0$ will win in $\mathsf{AugTourn}$ for item~$j$.
However, conditioned on $i_0$ losing item~$j$ at time $t_0$,
the induction hypothesis says that by time $t > t_0$,
$i_0$ would have claimed ownership of some item with probability~$\frac{t - t_0}{1 - t_0}$.

Hence, to summarize the case when item $j$ is fully consumed before time $t$,
the probability that $i_0$ will have claimed ownership of some item by~$t$ is:

$\frac{t_0 - t_1}{1 - t_1} + (1 - \frac{t_0 - t_1}{1 - t_1}) \cdot \frac{t - t_0}{1 - t_0}
= \frac{t - t_1}{1 - t_1}$.

This completes the induction proof.
  \end{proof}

\begin{lemma}[Uniform Dominance]
\label{lemma:u_dom}
The protocol $\mathsf{OnlinePSVar}$ achieves uniform dominance.
\end{lemma}

\begin{proof}
It suffices to show that for all $\ell \in [n]$,
the probability that an honest player~$i_0$
will receive an item among its top~$\ell$ favorite
items is at least $\frac{\ell}{n}$,
no matter whether the other players are truthful or honest.

Observe that the sum of consumption rates
over all players is at most $n$ (when a corrupted player
is detected, its rate is set to 0).
Therefore, before $t = \frac{\ell}{n}$,
at least one of player~$i_0$'s top $\ell$ items has not been totally consumed.
This means that before time~$t$, player~$i_0$ is consuming
only among its top~$\ell$ items, and it is obvious
that a player can only claim ownership of an item that it has attempted to consume.

Hence, by Lemma~\ref{lemma:ownership} with $t_0 = 0$,
the probability that player~$i_0$ has claimed ownership
of one of its top~$\ell$ items by time $t = \frac{\ell}{n}$
is $\frac{\ell}{n}$.
Finally, consider the natural coupling between
the interrupted process at time~$t$
and the original $\mathsf{OnlinePSVar}$ process
(achieved by coupling the Bernoulli process at interruption with the corresponding
$\mathsf{AugTourn}$)
such that any item whose ownership is claimed by~$i_0$ at time~$t$
is also received by~$i_0$ in~$\mathsf{OnlinePSVar}$.  This completes the proof.
  \end{proof}

\subsection{$\mathsf{OnlinePSVar}$ is not Maximin Secure nor Strongly Truthful}

We will give two examples to show that $\mathsf{OnlinePSVar}$
is neither maximin secure nor strongly truthful.

\noindent \textbf{$\mathsf{OnlinePSVar}$ is not Maximin Secure.}
To show that
$\mathsf{OnlinePSVar}$ cannot achieve maximin security even against fail-stop adversary, we first illustrate the idea of constructing the counter-example for showing OnlinePSVar is not maximin secure. Suppose there are two sets of players $X_1, X_2$ with large enough $|X_1|$ and $|X_2|$,  and three special players, $i_1, i_2, i_3$. The following is the profile regarding the three special players with unimportant items omitted:

$$m_3 \succ_1 \cdots \succ_1 \cdots \succ_1 \cdots$$
$$m_1 \succ_2 m_2 \succ_2 m_3 \succ_2 \cdots$$
$$m_1 \succ_3 m_2 \succ_3 m_3 \succ_3 \cdots$$

We assume players in $X_1$ like $m_1$ the most and will compete for $m_1$ at time $0$, and later they will not interfere with items $m_2$ and $m_3$. For  players in $X_2$, we assume that they will compete for $m_2$ at time $\frac{1}{2}$, but they will not interfere with $m_1$ and $m_3$ at all. We have two observations.

\begin{enumerate}
\item We consider the game with players $X_1\cup\{ i_1, i_2, i_3\}$. In this case, at time $0$, $i_2$, $i_3$ and players in $X_1$ will compete for $m_1$, and it will be finished in a short time. Since $i_2$ and $i_3$ have low chances of getting $m_1$, they will then compete for $m_2$ and finish this round at time $\approx \frac{1}{2}$. As a result, one of the players will get $m_2$ and the other will join in competing for $m_3$ with $i_1$ at time $\approx \frac{1}{2}$.

\item We consider the game with players $X_2\cup\{i_1, i_2, i_3\}$ In this case, one of $i_2$ and $i_3$ will get $m_1$ and the other (namely, $i_2$) will start to compete for $m_2$ at time $\frac{1}{2}$. Note that players in $X_2$ will join as well and item $m_2$ will be finished in a short time. Since $i_2$ has low chance for getting $m_2$, it will join in competing $m_3$ with $i_1$ at time $\approx \frac{1}{2}$.
\end{enumerate}

The above observations illustrate a fact that, when one of $X_1$ or $X_2$ is absent, the probability for $i_1$ getting $m_3$ in these two cases could be arbitrarily close. Hence, we consider a sequence of games constructed by replacing participants involved in the game: we start from the game described in observation $1$, and gradually shift to the one in observation $2$ by adding players from $X_2$ or removing players from $X_1$. Then, there exist two consecutive games, where, the probability of $i_1$ getting $m_3$ declines after removing some player from $X_1$, or increases after adding some player from $X_2$. This gives the intuition for the following counter-example.

\begin{example}[$\mathsf{OnlinePSVar}$ Not Maximin Secure]
Consider the preference profile given by the following matrix,
where each row corresponds to a player and each entry~$(i,j)$
contains the index of the $j$-th favorite item for player~$i$.

\setcounter{MaxMatrixCols}{20}

\begin{center}
$\left[ \begin{matrix}
3 & 1 & 2 & 4 & 5 & 6 & 7 & 8 & 9 & 10 & 11 & 12 & 13 & 14 \\
{1} &    {2} &    {3} &    {4} &    {5} &    {6} &    {7} &    {8} &    {9} &    {10} &  {11} &    {12} &    {13} &    {14} \\
{1} &    {2} &    {3} &    {4} &    {5} &    {6} &    {7} &    {8} &    {9} &    {10} &    {11} &    {12} &    {13} &    {14} \\
{1} &    {14} &    {13} &    {12} &    {11} &    {10} &    {9} &    {8} &    {7} &    {6} &  {5} &    {4} &    {2} &    {3} \\
{4} &    {2} &    {14} &    {13} &    {12} &    {11} &    {10} &    {9} &    {8} &    {7} &  {6} &    {5} &    {1} &    {3} \\
{4} &    {2} &    {14} &    {13} &    {12} &    {11} &    {10} &    {9} &    {8} &    {7} &  {6} &    {5} &    {1} &    {3} \\
{5} &    {2} &    {14} &    {13} &    {12} &    {11} &    {10} &    {9} &    {8} &    {7} &  {6} &    {4} &    {1} &    {3} \\
{5} &    {2} &    {14} &    {13} &    {12} &    {11} &    {10} &    {9} &    {8} &    {7} &  {6} &    {4} &    {1} &    {3} \\
{6} &    {2} &    {14} &    {13} &    {12} &    {11} &    {10} &    {9} &    {8} &    {7} &  {5} &    {4} &    {1} &    {3} \\
{6} &    {2} &    {14} &    {13} &    {12} &    {11} &    {10} &    {9} &    {8} &    {7} &  {5} &    {4} &    {1} &    {3} \\
{7} &    {2} &    {14} &    {13} &    {12} &    {11} &    {10} &    {9} &    {8} &    {6} &  {5} &    {4} &    {1} &    {3} \\
{7} &    {2} &    {14} &    {13} &    {12} &    {11} &    {10} &    {9} &    {8} &    {6} &  {5} &    {4} &    {1} &    {3} \\
{8} &    {2} &    {14} &    {13} &    {12} &    {11} &    {10} &    {9} &    {7} &    {6} &  {5} &    {4} &    {1} &    {3} \\
{8} &    {2} &    {14} &    {13} &    {12} &    {11} &    {10} &    {9} &    {7} &    {6} &  {5} &    {4} &    {1} &    {3}
\end{matrix} \right]$.
\end{center}

The player $4$ acts as $X_1$ and players $5,\dots, 14$ act as $X_2$ in the discussion above. Using a program, we can use the brute force approach to compute
the probability that a player wins a certain item.
 When all players act honestly, player~1 will get its favorite item $m_3$ with probability $\frac{1132927}{1499784} \approx 0.7553$.
However, if player~4 aborts at the beginning, then agent 1 gets $m_3$ with probability $\frac{77}{102} \approx 0.7549$, which violates maximin security.

\end{example}

\noindent \textbf{$\mathsf{OnlinePSVar}$ is not Strongly Truthful.}
The following example~\ref{ex:truthful} is a counter-example to show that $\mathsf{OnlinePSVar}$ is not strongly truthful.

\begin{example}[$\mathsf{OnlinePSVar}$ Not Strongly Truthful]
	\label{ex:truthful}
	Consider $n = 4$ players with the following true preferences:
	$$m_1 \succ_1 m_2 \succ_1 m_3 \succ_1 m_4$$
	$$m_1 \succ_2 m_2 \succ_2 m_3 \succ_2 m_4$$
	$$m_2 \succ_3 m_3 \succ_3 m_4 \succ_3 m_1$$
	$$m_2 \succ_4 m_3 \succ_4 m_4 \succ_4 m_1$$
	
	In
	$\mathsf{OnlinePSVar}$,
	player~2 obtains one of its top two items $\{m_1, m_2\}$ with probability~$\frac{1}{2}$.
	We show that player~2 can increase this probability by lying about its preference as:
	
	$$m_2 \succ m_1 \succ m_3 \succ m_4.$$
	
	At time $t = \frac{1}{3}$, item~$m_2$ is totally consumed.
	
	If player~2 loses the tournament for item~$m_2$,
	then it will compete with player~1 for item~$m_1$ starting at time $t = \frac{1}{3}$.
	In $\mathsf{OnlinePS}$, the rate of player~2 remains 1, while
	in $\mathsf{OnlinePSVar}$, its rate is increased to 1.5.  Hence,
	it suffices to do the calculation for the former case,
	in which item~$m_1$ will be totally consumed at time $t' = \frac{2}{3}$,
	when player~2 will get $\frac{1}{3}$ fraction of $m_1$.
	
	Hence, by lying, the probability that player~2 
	obtains either $m_1$ or $m_2$ is: $\frac{1}{3} + (1 - \frac{1}{3}) \cdot \frac{1}{3} = \frac{5}{9}$, which is larger than before.
\end{example}

\section{Conclusion}
\label{sec:conclusion}

We have considered the game-theoretic notion of maximin security
for protocols solving the ordinal assignment problem,
where randomness is necessary to achieve the fairness notion
of equal treatment.

Our major contribution is the impossibility result
that shows no maximin secure protocol can satisfy both
strong equal treatment and ordinal efficiency,
thereby also excluding the possibility of any maximin secure
protocol that realizes the well-known \PS mechanism.
However, the problem of whether there exists a maximin secure
protocol that realizes \RP is still open.  In general,
the following questions are interesting future directions.

\begin{compactitem}
\item Does there exist a maximin secure protocol that
achieves both strong equal treatment and uniform dominance?

\item Does there exist a maximin secure protocol that
achieves both strong equal treatment and truthfulness?
\end{compactitem}

{
\bibliography{reference,assignment,committee,other,byz}
\bibliographystyle{alpha}
}



\end{document}